\documentclass[12pt]{iopart}

\usepackage{epsfig}

\usepackage{iopams}

\begin{document}

\title[A cosmological model in Weyl-Cartan spacetime I]{A cosmological model in Weyl-Cartan spacetime:\quad I. Field equations and solutions}
\author{Dirk Puetzfeld}
\address{Institute for Theoretical Physics, University of Cologne,
  50923 K\"oln, Germany}
\ead{dp@thp.uni-koeln.de}

\begin{abstract}
In this first article of a series on alternative cosmological models
we present an extended version of a cosmological model in Weyl-Cartan
spacetime. The new model can be viewed as a generalization of a model
developed earlier jointly with Tresguerres. Within this model the
non-Riemannian quantities, i.e.\ torsion $T^{\alpha}$ and nonmetricity $Q_{\alpha \beta}$, are proportional to
the Weyl 1-form. The hypermomentum $\Delta_{\alpha \beta}$ depends on our ansatz for the nonmetricity and vice versa. We derive the explicit form of the field
equations for different cases and provide solutions for a broad class of
parameters. We demonstrate that it is possible to construct models in which the
non-Riemannian quantities die out with time. We show how our model fits into the more general framework of metric-affine gravity (MAG).  
\end{abstract}

\submitto{\CQG}
\maketitle

\section{Introduction}

At the moment cosmology is one of the fastest changing fields in physics. This
fact might, on the one hand, be ascribed to the vast amount of new
observational data (cf \cite{Perlmutter,Schmidt,Jaffe} e.g.), on the other hand there are
still fundamental open questions within what is nowadays called the
cosmological standard model \cite{KolbTurner,Peacock}: Where does the
inflaton field come from? Is there a something like the cosmological constant
$\lambda$ which contributes to the dark energy etc.? 

From a theoretical viewpoint one might divide efforts today within cosmology
into two broad subclasses. Firstly, we have models which extend the standard model to a
certain amount, inflation \cite{Guth1,Linde1}, e.g., can be viewed as an add-on
for the classical Friedman-Lema\^{\i}tre-Robertson-Walker (FLRW)
model. All of these models have in common that they do not affect the
structure of spacetime itself, i.e., they are still bound to a four-dimensional
Riemannian spacetime and, in addition, do not modify the underlying gravity
theory, i.e., General Relativity (GR). Secondly, we have models which are no longer tied to a four-dimensional Riemannian spacetime and might
also modify the underlying gravity theory. One example for the latter is the
so-called metric-affine gauge theory of gravity (MAG), as proposed by Hehl \etal
in \cite{PhysRep}. Within this gauge theoretical formulation of a gravity
theory there are new geometrical quantities, torsion $T^\alpha:=D \vartheta^\alpha$
and nonmetricity $Q_{\alpha \beta}:=-Dg_{\alpha\beta}$, which liberate spacetime from its
Riemannian structure. In this paper we are going to work within this more general
framework of a post-Riemannian gravity theory. 

The main reason to consider
more general structures within cosmology is the idea that new geometrical
quantities might shed light on the problems of the cosmological standard model,
e.g.\ provide an explanation for the rather artificial introduction of an
additional scalar field, like the inflaton field. The new quantities couple to
the spin, shear, and dilation currents of matter, which are supposed to come
into play at high energy densities, i.e.\ at early stages of the universe \cite{HehlHeyde,Bauerle}. Another open question, which
might also be attacked, is the origin of the large amount of dark energy,
predicted by recent supernova observations. A model which deals with a new
kind of dark matter interaction within a post-Riemannian theory has been
proposed by Tucker and Wang in \cite{TuckerDark}.       

Within this paper, we confine ourselves to a Weyl-Cartan spacetime. This type
of spacetime can be viewed as a special case of the more general metric-affine
framework, in which the tracefree part of the nonmetricity $Q_{\alpha \beta} \hspace{-0.8cm} \nearrow
\hspace{0.3cm}$ vanishes by definition. The reason
to consider this kind of restriction is twofold. Firstly, computations
are more feasible in a spacetime which is not endowed with the full MAG symmetries. Secondly, the Weyl-Cartan spacetime is, unlike
the Riemann-Cartan or the Weyl spacetime, still able to carry both of the new
field strengths nonmetricity and torsion. Let us note
that the metric-affine framework incorporates all of the above mentioned types
of spacetimes. The Einstein-Cartan theory, which is formulated in a
Riemann-Cartan spacetime, represents a viable gravity
theory with torsion. By switching off all non-Riemannian quantities in MAG we
arrive at GR. 

As becomes clear from the title, this paper stands at the beginning of series of articles
on cosmological models in alternative gravity theories. In this first article
we derive the field equations for an enhanced gauge Lagrangian and look for
solutions of these equations. Thereby we extend earlier joint work with Tresguerres \cite{PuetzTres} and lay the foundation for the second article in
this series, which will deal with the observational consequences of this new
cosmological model.  

The plan of this paper is as follows. In \sref{TRESGUERRES_KAPITEL} we
derive the field equations and Noether identities for a new gauge Lagrangian
on a formal level. After that we make use of computer algebra and provide the
explicit form of the field equations and Noether identities for a rather
general choice of the Weyl 1-form in
\sref{FIELD_NOETHER_ARB_SECTION}. In \sref{SPECIAL_CASE_SECTION} we restrict
our considerations to a special case, which
leads to a more manageable set of field equations. In
\sref{special_case_xi_=_zeta_solutions_section} we look for exact solutions of
these equations. We will draw our conclusion in
\sref{CONCLUSION_SECTION} and present some plots for one branch of our model.  In \ref{MAG_KAPITEL} and
\ref{WEYL_CARTAN_KAPITEL} we provide a short introduction into MAG and the
Weyl-Cartan spacetime. Additionally, we show in \ref{WEYL_CARTAN_KAPITEL} how our model fits into the general framework of MAG
as proposed in \cite{PhysRep}. In \ref{UNITS_SECTION}, we provide an
overview over the units used throughout the preceding sections. Note that we make extensive use of differential
forms within this paper. A short compilation of symbols used within the paper can be found in
\ref{OPERATIONS_SECTION}, for a more rigorous treatment the reader should
consult Appendix A of \cite{PhysRep}.

\section{Lagrangian, gauge, and matter currents \label{TRESGUERRES_KAPITEL}}

In \cite{PuetzTres} we considered the following gauge Lagrangian 
\begin{equation}
V_{\texttt{\tiny \rm old}}=\frac{\chi }{2\kappa }R_{\alpha }{}^{\beta }\wedge \,\eta
_{\beta }{}^{\alpha }+\sum_{I=1}^{6}a_{I}\,^{\left( I\right) }W_{\alpha
}{}^{\beta }\wedge \,^{\star }R_{\beta }{}^{\alpha }+b \, Z_{\alpha \beta
}\wedge \,^{\star }R^{\beta \alpha }.  \label{lag_old}
\end{equation}
Where $R_{\alpha \beta}=W_{\alpha \beta}+Z_{\alpha \beta} =$ antisymmetric +
symmetric part of the curvature, and $\eta_{\alpha
  \beta}:=\,^{\star}\left(\vartheta_\alpha \wedge \vartheta_\beta
\right)$. Numbers in parentheses in front of quantities correspond to the
irreducible decompositions performed in \cite{McCrea}. 

Since we are interested in more general gauge Lagrangians (cf equation
(\ref{general_v_mag}) for a very general one proposed in MAG), we are going to extend (\ref{lag_old}). We will perform our calculations on the basis of a Weyl-Cartan
spacetime, i.e.\ a spacetime in which the tracefree part of the nonmetricity
$Q_{\alpha \beta}$ vanishes, a short introduction into Weyl-Cartan spacetime is given in
\ref{WEYL_CARTAN_KAPITEL}. The most obvious extension of (\ref{lag_old}) is given by 
\begin{equation}
V_{\texttt{\tiny \rm 1}}=\sum_{I=1}^{4}c_{I}\,\,^{\left( I\right) }Q_{\alpha \beta
}\,\wedge \,^{\star }Q^{\beta \alpha }.  \label{quadratic_nonmet_term}
\end{equation}
Since in a Weyl-Cartan space the nonmetricity is reduced to its trace part,
i.e.\ $Q_{\alpha \beta }=g_{\alpha \beta }Q=\frac{1}{4}g_{\alpha \beta
}Q^{\gamma }{}_{\gamma }=\,^{\left( 4\right) }Q_{\alpha \beta }$, see equation
(\ref{torsion_in_weyl_cartan_spacetime}), equation (\ref{quadratic_nonmet_term}) now reads 
\begin{equation}
V_{\texttt{\tiny \rm 1}}=c_{4}\,\,^{\left( 4\right) }Q_{\alpha \beta }\,\wedge \,^{\star
}\,^{\left( 4\right) }Q^{\beta \alpha } =: c\,Q_{\alpha \beta }\wedge \,^{\star }Q^{\beta \alpha }.
\label{short_new_lagrangian}
\end{equation}
Our new Lagrangian now reads 
\begin{eqnarray}
V&=&V_{\texttt{\tiny \rm old}}+V_{\texttt{\tiny \rm 1}} \label{Langrangian_new_symbolic}\\
&=&\texttt{\rm Einstein-Hilbert} + \texttt{\rm quadratic rotational curvature}  \nonumber \\
&&\texttt{\rm + quadratic strain curvature + quadratic nonmetricity.}
\label{lag_new}
\end{eqnarray}
In contrast to \cite{Minkevich4} we included an explicit nonmetricity term in
our Lagrangian. Note that we have the arbitrary constants $\chi ,$ $a_{I=1\dots
  6},$ $b,$ $c, $ and the {\it weak} gravity coupling constant $\kappa $. The Lagrangian in (\ref{lag_new}) can be viewed as another step
towards a better understanding of the full MAG Lagrangian as displayed in
(\ref{general_v_mag}). Since a treatment of the full Lagrangian is
computationally not feasible at the moment it is necessary to successively study the impact
of additional terms in the Lagrangian (a review of
Lagrangians used in MAG and exact solutions of the corresponding field
equations can be found in \cite{Exact2}). Together with the quadratic rotational
curvature and quadratic strain curvature terms, which were already included in our previous work
\cite{PuetzTres}, we now have an additional post-Riemannian piece in form of a
quadratic nonmetricity term which enhances the usual Einstein-Hilbert
Lagrangian commonly used in general relativistic cosmological models. Note
that our Lagrangian in (\ref{lag_new}) does not include a term with the usual cosmological constant. As we will show in section \ref{SPECIAL_CASE_SECTION} our ansatz
in (\ref{lag_new}) gives rise to an additional constant which, on the level of the
field equations, will play the same role as the cosmological constant in
the standard model. Hence we omit an explicit cosmological constant term at
this stage. From (\ref{lag_new}) we can derive the gauge field excitations. They read 
\begin{eqnarray}
M^{\alpha \beta } &=&-4c\,^{\star }Q^{\beta \alpha }=-c\,^{\star
}\left( g^{\beta \alpha }Q^{\gamma }{}_{\gamma }\right) ,  \label{excit_1} \\
H_{\alpha } &=&0,  \label{excit_3} \\
H^{\alpha }{}_{\beta } &=&-\frac{\chi }{2\kappa }\eta _{\beta }{}^{\alpha
}-2\sum_{I=1}^{6}a_{I}\,^{\star (I)}W_{\beta }{}^{\alpha }-\frac{b}{2}\delta
_{\beta }^{\alpha }\,^{\star }R^{\gamma }{}_{\gamma }.  \label{excit_2}
\end{eqnarray}
The canonical gauge energy-momentum is given by 
\begin{eqnarray}
E_{\alpha } &=&e_{\alpha }\rfloor V+\left( e_{\alpha }\rfloor R_{\beta
}{}^{\gamma }\right) \wedge H^{\beta }{}_{\gamma }+\left( e_{\alpha }\rfloor
T^{\beta }\right) \wedge H_{\beta }+\frac{1}{2}\left( e_{\alpha }\rfloor
Q_{\beta \gamma }\right) \wedge M^{\beta \gamma }  \nonumber \\
&=&e_{\alpha }\rfloor V+\left( e_{\alpha }\rfloor R_{\beta }{}^{\gamma
}\right) \wedge H^{\beta }{}_{\gamma }+\frac{1}{2}\left( e_{\alpha }\rfloor
Q_{\beta \gamma }\right) \wedge M^{\beta \gamma }.
\end{eqnarray}
In contrast to \cite{PuetzTres}, we now have a non-vanishing gauge
hypermomentum \footnote{Additional assumptions are marked with an ''$A$''. Note that we mark new relations, new with respect to \cite{PuetzTres}, with
an ''$N$''.} 
\begin{eqnarray}
E^{\alpha }{}_{\beta } &=&-\vartheta ^{\alpha }\wedge H_{\beta }-g_{\beta
\gamma }M^{\alpha \gamma }  \nonumber \\
&\stackrel{N}{=}&4\, c\, g_{\beta \gamma }\,^{\star }Q^{\gamma \alpha }=\, c \,g_{\beta \gamma
}\,^{\star }\left( g^{\gamma \alpha }Q^{\nu }{}_{\nu }\right) .
\label{hypermomentum}
\end{eqnarray}
The field equations now turn into 
\begin{eqnarray}
-E_{\alpha } &=&\Sigma _{\alpha },  \label{field1} \\
dH^{\alpha }{}_{\alpha }-E^{\alpha }{}_{\alpha } &\stackrel{N}{=}&\Delta ,  \label{field2_trace} \\
g_{\gamma \lbrack \alpha }DH^{\gamma }{}_{\beta ]}-E_{[\alpha \beta ]} &\stackrel{N}{=}&\tau _{\alpha \beta }.  \label{field2_antisymmetric}
\end{eqnarray}
Note that in eqs.\ (\ref{field2_trace}) and (\ref{field2_antisymmetric}) we
decomposed the second field equation into its trace and antisymmetric part
(cf \ref{WEYL_CARTAN_KAPITEL}). Since we are interested in the behaviour
induced by the new part in (\ref{lag_new}), we will confine ourselves to a
{\em non-massive medium without spin}, i.e.\ $\tau _{\alpha \beta } \stackrel{A}{=}0$. Thus, eq.\ (\ref{field2_antisymmetric}) turns into 
\begin{equation}
g_{\gamma \lbrack \alpha }DH^{\gamma }{}_{\beta ]}-E_{[\alpha \beta ]}\stackrel{N}{=}0.  \label{field2_antisymmtetric_vanishing_spin_current}
\end{equation}
Because we have not specified a matter Lagrangian, we have to take into
account the Noether identities (cf \ref{MAG_KAPITEL} and \ref{WEYL_CARTAN_KAPITEL}), i.e.\ 
\begin{eqnarray}
D\Sigma _{\alpha } &=&\left( e_{\alpha }\rfloor T^{\beta }\right) \wedge
\Sigma _{\beta }-\frac{1}{2}\left( e_{\alpha }\rfloor Q\right) \sigma
^{\beta }{}_{\beta } +\left( e_{\alpha }\rfloor R_{[\beta \gamma ]}\right)
\wedge \tau ^{\beta \gamma } \nonumber \\ && +\frac{1}{4}\left( e_{\alpha }\rfloor R\right)
\wedge \Delta ,  \label{noether1} \\
\sigma _{\alpha \beta } &=&\frac{1}{4}g_{\alpha \beta }d\Delta +\vartheta
_{(\alpha }\wedge \Sigma _{\beta )},  \label{noether2_1} \\
0 &=&\vartheta _{\lbrack \alpha }\wedge \Sigma _{\beta ]}.
\label{noether2_2}
\end{eqnarray}
We can rewrite eq.\ (\ref{noether2_1}) by using (\ref{field2_trace}) 
\begin{equation}
\sigma _{\alpha \beta }\stackrel{N}{=}-\frac{1}{4}g_{\alpha \beta }\,dE^{\gamma }{}_{\gamma }+\vartheta _{(\alpha
}\wedge \Sigma _{\beta )}.  \label{metric_stress_energy_relation}
\end{equation}
Note that (\ref{noether2_1})-(\ref{noether2_2}) represent the decomposed
second Noether identity in case of a vanishing spin current. With (\ref
{noether2_2}), eq.\ (\ref{noether1}) turns into 
\begin{eqnarray}
D\Sigma _{\alpha }&\stackrel{N}{=}&\left( e_{\alpha }\rfloor T^{\beta }\right) \wedge \Sigma _{\beta }-\frac{1}{2}\left( e_{\alpha }\rfloor Q\right) \vartheta ^{\beta }\wedge \Sigma
_{\beta }+\frac{1}{8}\left( e_{\alpha }\rfloor Q\right) \,g^{\beta
}{}_{\beta }\,dE^{\gamma }{}_{\gamma }\nonumber \\
&&+\frac{1}{4}\left( e_{\alpha }\rfloor
R\right) \wedge \Delta .  \label{noether1_simp1}
\end{eqnarray}
Thus, we have to solve (\ref{field1}), (\ref{field2_trace}), (\ref
{field2_antisymmtetric_vanishing_spin_current}), and (\ref{noether2_2})--(\ref
{noether1_simp1}) in order to obtain a solution for our model proposed in (%
\ref{lag_new}). Now let us investigate the matter sources of our model. For
a vanishing spin current, the hypermomentum $\Delta _{\alpha \beta }$ becomes
proportional to its trace part, i.e.\ the dilation current cf \eref{Weyl_Cartan_hypermomentum}
\begin{equation}
\Delta _{\alpha \beta }=\frac{1}{4}g_{\alpha \beta }\Delta ^{\gamma
}{}_{\gamma }.  \label{general_hypermomentum_for_vanishing_spin_current}
\end{equation}
The trace part of the second field equation (\ref{field2_trace}) yields 
\begin{equation}
\Delta _{\alpha \beta }=\frac{1}{4}g_{\alpha \beta }\left( dH^{\gamma
}{}_{\gamma }-E^{\gamma }{}_{\gamma }\right) .  \label{dilation_current}
\end{equation}
In contrast to \cite{PuetzTres}, the dilation current is no longer a
conserved quantity since $d\Delta_{\alpha \beta }\not=0$. From (\ref{excit_2}), and (\ref{hypermomentum}) we
can infer that 
\begin{equation}
\Delta _{\alpha \beta }=-\frac{1}{4}g_{\alpha \beta }\left( 2\,b\,d\,^{\star
}R^{\gamma }{}_{\gamma }+c\,g_{\gamma \mu }\,^{\star }\left( g^{\mu \gamma
}Q^{\nu }{}_{\nu }\right) \right) .  \label{dilation_current_explicit}
\end{equation}
Because of \eref{strain_curvature_in_weyl_cartan_general}, i.e.\ $R^{\gamma }{}_{\gamma }\sim dQ$,
this equation turns into 
\begin{equation}
\Delta _{\alpha \beta }\stackrel{N}{=}-\frac{1}{4}g_{\alpha \beta }\left( \,b\,d\,^{\star }dQ^{\gamma }{}_{\gamma
}+c\,g_{\gamma \mu }\,^{\star }\left( g^{\mu \gamma }Q^{\nu }{}_{\nu
}\right) \right) .  \label{hypermomentum_with_inserted_strain_curvature}
\end{equation}
Thus, for our ansatz the hypermomentum $\Delta _{\alpha \beta }$ depends on the nonmetricity and vice versa. Note that the second term in (\ref
{hypermomentum_with_inserted_strain_curvature}) depends on the coupling
constant introduced in eq.\ (\ref{excit_1}). Now let us specify the remaining
quantities in our model. \Eref{noether2_2} forces the
components of energy-momentum 3-form to be symmetric, thus we choose\footnote{%
Here we made use of $\eta ^{\alpha }:=\,^{\star }\vartheta ^{\alpha }$.} 
\begin{equation}
\Sigma _{\alpha }\stackrel{A}{=}\Sigma _{\alpha \beta }\,\eta ^{\beta }\quad \quad \texttt{\rm , with}\quad \Sigma
_{\alpha \beta }=\texttt{\rm diag}\left( \mu (t),p_{r}(t),p_{t}(t),p_{t}(t)\right) 
\texttt{\rm .}  \label{energy_momentum_ansatz}
\end{equation}
Subsequently, we can calculate the metric stress-energy $\sigma _{\alpha
\beta }$ from eq.\ (\ref{metric_stress_energy_relation}) 
\begin{equation}
\sigma _{\alpha \beta }=-\frac{1}{4}g_{\alpha \beta }\,\,d\left(
c\,\,g_{\gamma \mu }\,^{\star }\left( g^{\mu \gamma }Q^{\nu }{}_{\nu
}\right) \right) +\vartheta _{(\alpha }\wedge \left( \Sigma _{\beta )\gamma
}\,\eta ^{\gamma }\right) .  \label{metric_stress_energy_explicit}
\end{equation}
Again we obtained a quantity which depends on the Weyl 1-form, i.e.\ the
trace of the nonmetricity. Since we want to compare our model with the
cosmological standard model, we take the Robertson-Walker line element as
starting point of our considerations 
\begin{equation}
\vartheta ^{\hat{0}}=dt,\quad \vartheta ^{\hat{1}}=\frac{S(t)}{\sqrt{1-kr^{2}%
}}dr,\quad \vartheta ^{\hat{2}}=S(t)\,r\,d\theta ,\quad \vartheta ^{\hat{3}%
}=S(t)\,r\sin \theta \, d\phi ,  \label{robertson_walker_coframe}
\end{equation}
with 
\begin{equation}
ds^{2}\stackrel{A}{=}\vartheta ^{\hat{0}}\otimes \vartheta ^{\hat{0}}-\vartheta ^{\hat{1}}\otimes
\vartheta ^{\hat{1}}-\vartheta ^{\hat{2}}\otimes \vartheta ^{\hat{2}%
}-\vartheta ^{\hat{3}}\otimes \vartheta ^{\hat{3}}.
\label{robertson_walker_line_element}
\end{equation}
As usual, $S(t)$ denotes the cosmic scale factor and $k=-1,0,$ or $1$ determines
whether the three-dimensional spatial sections of spacetime are of constant
negative, vanishing, or positive Riemannian curvature. Following the model proposed in \cite{PuetzTres}, we will choose
the torsion to be proportional to its vector piece $T^{\alpha }\sim
\,^{(2)}T^{\alpha }$ and relate it to the Weyl 1-form as follows 
\begin{equation}
T^{\alpha }\stackrel{A}{=}\frac{1}{2}Q\wedge \vartheta ^{\alpha }.  \label{ansatz_torsion_1}
\end{equation}
The only thing missing for setting up the field equations is a
proper ansatz for the Weyl 1-form $Q$. In \cite{PuetzTres} we were able to
derive $Q$ from an ansatz for the potential of the hypermomentum $\Delta$, the so called polarization 2-form $P$. Here we will adopt a
slightly different point of view. Since we are interested in the impact of
different choices of the non-Riemannian quantity $Q$ on cosmology, we will
directly prescribe it in the following. Besides of the fact that we gain direct control
of the post-Riemannian features of our model, we circumvent the question which
type of matter might generate the corresponding hypermomentum. This question
and the investigation of models with a more sophisticated matter model, like
the hyperfluid of Obukhov \etal \cite{Obukhov2}, will be postponed to later
articles. Let us note that our ansatz in equation (\ref{energy_momentum_ansatz}) is in general not compatible with the energy-momentum obtained in
(\cite{Obukhov2}, eq.\ (3.28)). Both quantities are only equal in special
cases like the one we will investigate in section
\ref{SPECIAL_CASE_SECTION}. Since we do not prescribe a matter Lagrangian and
use the Noether identities as constraints on the matter variables, our
approach could be termed {\it phenomenological} as suggested in the first part
of \cite{Obukhov2}.

\section{Field equations and Noether identities\label{FIELD_NOETHER_ARB_SECTION}}

In this section we will derive the field equations and Noether identities
resulting from specific choices of the 1-form $Q$ which controls nearly
every feature of our model. We start with a rather general form of $Q$,
namely 
\begin{equation}
Q=\frac{\xi (t,r)}{S(t)}\,\vartheta ^{\hat{0}},
\label{specific_ansatz_for_nonemtricity_trace_no_1}
\end{equation}
where $\xi (t,r)$ denotes an arbitrary function\footnote{%
Note that this function is \textit{not} identical with the one used in (\cite
{PuetzTres}, eq.\ (24)). It has a slightly different meaning since we use it here
directly in our ansatz for the nonmetricity.} of the radial and the time
coordinate, and $S(t)$ represents the cosmic scale factor of (\ref
{robertson_walker_coframe}). With the help of computer algebra we find that
the field equations (\ref{field1}), (\ref{field2_trace}), and (\ref
{field2_antisymmtetric_vanishing_spin_current}) yield a set of four equations. In order to compare these new field equations with the ones derived in \cite{PuetzTres} (cf\ eqs.\ (40)-(43) therein) we write them as follows: 
\begin{eqnarray}
&&\chi \left( \left( \frac{\dot{S}}{S}\right) ^{2}+\frac{k}{S^{2}}\right)
-\left( a_{4}+a_{6}\right) \kappa \left( \left( \frac{\ddot{S}}{S}\right)
^{2}-\left[ \left( \frac{\dot{S}}{S}\right) ^{2}+\frac{k}{S^{2}}\right]
^{2}\right)  \nonumber \\
&& \quad \quad \quad = \frac{\kappa }{3}\left( \mu -4c\left( \frac{\xi }{S}\right) ^{2}+b\left(
1-kr^{2}\right) \frac{\xi _{,r}^{2}}{S^{4}}\right) ,  \label{nexp_field_1} \\
&&\chi \left( 2\frac{\ddot{S}}{S}+\left( \frac{\dot{S}}{S}\right) ^{2}+\frac{%
k}{S^{2}}\right) +\left( a_{4}+a_{6}\right) \kappa \left( \left( \frac{\ddot{%
S}}{S}\right) ^{2}-\left[ \left( \frac{\dot{S}}{S}\right) ^{2}+\frac{k}{S^{2}%
}\right] ^{2}\right)  \nonumber \\
&&\quad \quad \quad = -\kappa \left( p_{r}-4c\left( \frac{\xi }{S}\right) ^{2}-b\left(
1-kr^{2}\right) \frac{\xi _{,r}^{2}}{S^{4}}\right) ,  \label{nexp_field_2} \\
&&\chi \left( 2\frac{\ddot{S}}{S}+\left( \frac{\dot{S}}{S}\right) ^{2}+\frac{%
k}{S^{2}}\right) +\left( a_{4}+a_{6}\right) \kappa \left( \left( \frac{\ddot{%
S}}{S}\right) ^{2}-\left[ \left( \frac{\dot{S}}{S}\right) ^{2}+\frac{k}{S^{2}%
}\right] ^{2}\right)  \nonumber \\
&&\quad \quad \quad = -\kappa \left( p_{t}-4c\left( \frac{\xi }{S}\right) ^{2}+b\left(
1-kr^{2}\right) \frac{\xi _{,r}^{2}}{S^{4}}\right) ,  \label{nexp_field_3} \\
&&\frac{d}{dt}\left( \frac{\ddot{S}}{S}+\left( \frac{\dot{S}}{S}\right) ^{2}+%
\frac{k}{S^{2}}\right) =0.  \label{nexp_field_4}
\end{eqnarray}
Since we have not specified a matter Lagrangian, we have to be careful with
the Noether identities (\ref{noether1})-(\ref{noether2_2}). Our ansatz (\ref
{specific_ansatz_for_nonemtricity_trace_no_1}) yields two equations 
\begin{eqnarray}
\hspace{-1cm} \dot{S}S^{3}\left( 3\mu +p_{r}+2p_{t}\right) +\dot{\mu}S^{4}-16\dot{S}cS\xi
^{2}+2\xi _{,rt}\xi _{,r}\, b\left( 1-kr^{2}\right) -8\xi _{,t} \, cS^{2}\xi &=&0,
\label{exp_noether_1} \\
\xi _{,rr}\xi _{,r}\, br\left( 1-kr^{2}\right) +\xi _{,r}^{2}\,b\left(
2-3kr^{2}\right) +4\xi _{,r}\, crS^{2}\xi +S^{4}\left( p_{t}-p_{r}\right) &=&0.
\label{exp_noether_2}
\end{eqnarray}
Note that in eq.\ (\ref{nexp_field_4}) we assumed that $a_{4}\not=-a_{6}$. As one realizes immediately, eqs.\ (\ref{nexp_field_1})-(\ref{nexp_field_4})
are very similar to the ones we obtained in (\cite{PuetzTres}, (40)-(43)).
There is only a change on the rhs, i.e.\ the matter side, of the above
equations in form of additional terms contributing to the pressure and
energy density. As we can see from eqs.\ (\ref{nexp_field_1})-(\ref
{nexp_field_3}), the terms proportional to $\xi _{,r}$ vanish if we make the
same assumptions as in \cite{PuetzTres}, i.e.\ $\xi (t,r)\rightarrow \xi (t)$%
. Apart from this feature, there is another, more subtle change in (\ref
{nexp_field_1})-(\ref{nexp_field_4}), i.e.\ a term of the order $\xi ^{2}$
controlled by the new coupling constant $c$ (cf\ eq.\ (\ref
{short_new_lagrangian}))\footnote{%
Formerly the term of the order $\xi ^{2}$ was controlled by the coupling
constant $b$ (cf\ eq.\ (\ref{lag_old})), and (\cite{PuetzTres}, (41)-(43)).}.
The Noether identities (\ref{exp_noether_1}) and (\ref{exp_noether_2}) can
be transformed to 
\begin{eqnarray}
\hspace{-1cm}\frac{\partial }{\partial t}\left( \mu S^{4}+\xi _{,r}^{2}\, b\left(
1-kr^{2}\right) -8c\left( S\xi \right) ^{2}\right) +4cS^{2}\frac{\partial
\xi ^{2}}{\partial t}=\frac{1}{4}\frac{dS^{4}}{dt}\left( \mu
-p_{r}-2p_{t}\right) ,  \label{nexp_noether_1} \\
p_{r}-p_{t}=\frac{2cr}{S^{2}}\frac{\partial \xi ^{2}}{\partial r}+\frac{b}{%
S^{4}}\left( \frac{r}{2}\left( 1-kr^{2}\right) \frac{\partial \xi _{,r}^{2}}{%
\partial r}+\left( 2-3kr^{2}\right) \xi _{,r}^{2}\right) .
\label{nexp_noether_2}
\end{eqnarray}
Comparison of (\ref{nexp_noether_2}) with (\cite{PuetzTres}, eq.\ (37)) yields a
more sophisticated relation between the radial and tangential stresses.

Let us now extract some more information from the field equations. Addition
of (\ref{nexp_field_1}) and (\ref{nexp_field_3}) yields 
\begin{equation}
\hspace{-1cm}2\chi \left( \frac{\ddot{S}}{S}+\left( \frac{\dot{S}}{S}\right) ^{2}+\frac{k%
}{S^{2}}\right) =\frac{\kappa }{3}\left( \mu -3p_{t}+8c\left( \frac{\xi }{S}%
\right) ^{2}-2b\left( 1-kr^{2}\right) \frac{\xi _{,r}^{2}}{S^{4}}\right).
\label{nfield1_+_nfield3}
\end{equation}
Subtracting (\ref{nexp_field_3}) from (\ref{nexp_field_1}) yields 
\begin{eqnarray}
&&2\chi \frac{\ddot{S}}{S}+2\kappa \left( a_{4}+a_{6}\right) \left( \left( 
\frac{\ddot{S}}{S}\right) ^{2}-\left[ \left( \frac{\dot{S}}{S}\right) ^{2}+%
\frac{k}{S^{2}}\right] ^{2}\right)  \nonumber \\
&&\quad\quad\quad =-\frac{\kappa }{3}\left( \mu +3p_{t}-16c\left( \frac{\xi }{S}\right)
^{2}+4b\left( 1-kr^{2}\right) \frac{\xi _{,r}^{2}}{S^{4}}\right) .
\label{nfield1_-_nfield3}
\end{eqnarray}
Let us now combine (\ref{nexp_field_4}) and (\ref{nfield1_+_nfield3}) 
\begin{eqnarray}
0 &&\stackrel{\texttt{\rm(\ref{nexp_field_4})}}{=}2\chi \frac{d}{dt}\left( \frac{\ddot{S}}{S}+\left( \frac{\dot{S}}{S}\right)
^{2}+\frac{k}{S^{2}}\right)  \nonumber \\
&&\stackrel{\texttt{\rm(\ref{nfield1_+_nfield3})}}{=}\frac{\kappa }{3}\frac{d}{dt}\left( \mu -3p_{t}+8c\left( \frac{\xi }{S}%
\right) ^{2}-2b\left( 1-kr^{2}\right) \frac{\xi _{,r}^{2}}{S^{4}}\right) .
\label{combined_field4_and_addition}
\end{eqnarray}
The trace of the energy-momentum reads 
\begin{eqnarray}
\hspace{-0.5cm}\Sigma ^{\gamma }{}_{\gamma } =&\mu -p_{r}-2p_{t}  \nonumber \\
\stackrel{\texttt{\rm(\ref{nexp_noether_2})}}{=}& \mu -3p_{t}-\frac{2cr}{S^{2}}\frac{\partial \xi ^{2}}{\partial r}+\frac{b}{S^{4}}\left( \frac{r}{2}\left( 1-kr^{2}\right) \frac{\partial \xi _{,r}^{2}}{\partial r}+\left( 2-3kr^{2}\right) \xi _{,r}^{2}\right).
\label{energy_momentum_trace_field_inserted}
\end{eqnarray}
Since we encountered a system of coupled PDEs, we will confine us to a
special case in the following in which the field equations turn into a set
of coupled ODEs. At this point we would like to note that the above situation is
reminiscent to the extensions of the classical FLRW model to anisotropic and
inhomogeneous metrical structures.
For completeness we list the surviving curvature pieces for the ansatz in
equation \eref{specific_ansatz_for_nonemtricity_trace_no_1}
\begin{eqnarray}
^{(4)}W^{\alpha \beta } &=&\frac{\ddot{S}S-\dot{S}^{2}-k}{2S^{2}} \, \vartheta
^{\alpha }\wedge \vartheta ^{\beta },
\label{EXTWEYL_ricsymf_for_general_ansatz} \\
^{(6)}W^{\alpha \beta } &=&\frac{\ddot{S}S+\dot{S}^{2}+k}{2S^{2}}\, \vartheta
^{\alpha }\wedge \vartheta ^{\beta },
\label{EXTWEYL_scalar_for_general_ansatz} \\
^{(4)}Z_{\hat{0}\hat{0}} &=&-\,^{(4)}Z_{\hat{1}\hat{1}}=\,-^{(4)}Z_{\hat{2}%
\hat{2}}=-\,^{(4)}Z_{\hat{3}\hat{3}}=-\frac{\xi _{,r}\sqrt{-kr^{2}+1}}{2S^{2}%
}\, \vartheta ^{\hat{0}}\wedge \vartheta ^{\hat{1}}.
\label{EXTWEYL_dilcurv_for_general_ansatz}
\end{eqnarray}

\section{Special case $\xi (t,r)\rightarrow \zeta (t)$ \label{SPECIAL_CASE_SECTION}}

In this section we will investigate the interesting special case in which $Q$,
cf\ eq.\ (\ref{specific_ansatz_for_nonemtricity_trace_no_1}), is given by a
closed 1-form, i.e.\ 
\begin{equation}
Q=\frac{\zeta (t)}{S(t)}\vartheta ^{\hat{0}}.
\label{specific_ansatz_for_nonemtricity_trace_no_2}
\end{equation}
The field equations are now given by
\begin{eqnarray}
&&\chi \left( \left( \frac{\dot{S}}{S}\right) ^{2}+\frac{k}{S^{2}}\right)
-\left( a_{4}+a_{6}\right) \kappa \left( \left( \frac{\ddot{S}}{S}\right)
^{2}-\left[ \left( \frac{\dot{S}}{S}\right) ^{2}+\frac{k}{S^{2}}\right]
^{2}\right)  \nonumber \\
&&\quad \quad \quad = \frac{\kappa }{3}\left( \mu -4c\left( \frac{\zeta }{S}\right) ^{2}\right)
,  \label{nfield_zeta_t_1} \\
&&\chi \left( 2\frac{\ddot{S}}{S}+\left( \frac{\dot{S}}{S}\right) ^{2}+\frac{%
k}{S^{2}}\right) +\left( a_{4}+a_{6}\right) \kappa \left( \left( \frac{\ddot{%
S}}{S}\right) ^{2}-\left[ \left( \frac{\dot{S}}{S}\right) ^{2}+\frac{k}{S^{2}%
}\right] ^{2}\right)  \nonumber \\
&&\quad \quad \quad = -\kappa \left( p_{r}-4c\left( \frac{\zeta }{S}\right) ^{2}\right) ,
\label{nfield_zeta_t_2} \\
&&\chi \left( 2\frac{\ddot{S}}{S}+\left( \frac{\dot{S}}{S}\right) ^{2}+\frac{%
k}{S^{2}}\right) +\left( a_{4}+a_{6}\right) \kappa \left( \left( \frac{\ddot{%
S}}{S}\right) ^{2}-\left[ \left( \frac{\dot{S}}{S}\right) ^{2}+\frac{k}{S^{2}%
}\right]^{2}\right)  \nonumber \\
&&\quad \quad \quad = -\kappa \left( p_{t}-4c\left( \frac{\zeta }{S}\right) ^{2}\right) ,
\label{nfield_zeta_t_3} \\
&&\frac{d}{dt}\left( \frac{\ddot{S}}{S}+\left( \frac{\dot{S}}{S}\right) ^{2}+%
\frac{k}{S^{2}}\right) =0.  \label{nfield_zeta_t_4}
\end{eqnarray}
Thus, the function $\zeta $ contributes to the energy density and pressure
in a similar way as the function $\xi $ in \cite{PuetzTres}. We note that
there is no additional contribution from the strain curvature in eqs.\ (\ref
{nfield_zeta_t_1})--(\ref{nexp_field_4}), i.e.\ no term controlled by the
coupling constant $b$ of our Lagrangian (cf\ eq.\ (\ref{lag_old})). This
behaviour is explained by the fact that the strain curvature vanishes
identically for closed 1-forms, like $Q$ from eq.\ (\ref
{specific_ansatz_for_nonemtricity_trace_no_2}), in a Weyl-Cartan spacetime.
The Noether identities now read:
\begin{eqnarray}
&&\frac{d}{dt}\left( \mu S^{4}-8c\left( S\zeta \right) ^{2}\right) +4cS^{2}%
\frac{d\zeta ^{2}}{dt}=\frac{1}{4}\frac{dS^{4}}{dt}\left( \mu
-p_{r}-2p_{t}\right) ,  \label{nnoether_zeta_t_1} \\
&&p_{r}-p_{t}=0.  \label{nnoether_zeta_t_2}
\end{eqnarray}
In contrast to (\ref{nexp_noether_2}), eq.\ (\ref{nnoether_zeta_t_2}) forces
the radial stress to be equal to the tangential stress. Addition of (\ref{nfield_zeta_t_1}%
) and (\ref{nfield_zeta_t_3}), i.e.\ eq.\ (\ref{nfield1_+_nfield3}), yields 
\begin{equation}
2\chi \left( \frac{\ddot{S}}{S}+\left( \frac{\dot{S}}{S}\right) ^{2}+\frac{k%
}{S^{2}}\right) =\frac{\kappa }{3}\left( \mu -3p_{r}+8c\left( \frac{\zeta }{S%
}\right) ^{2}\right) .  \label{nfield_zeta1_+_nfield_zeta_3}
\end{equation}
Subtracting (\ref{nfield_zeta_t_3}) from (\ref{nfield_zeta_t_1}) (cf\ eq.\ 
\ref{nfield1_-_nfield3}) yields 
\begin{eqnarray}
&&2\chi \frac{\ddot{S}}{S}+2\kappa \left( a_{4}+a_{6}\right) \left( \left( 
\frac{\ddot{S}}{S}\right) ^{2}-\left[ \left( \frac{\dot{S}}{S}\right) ^{2}+%
\frac{k}{S^{2}}\right] ^{2}\right)  \nonumber \\
&&\quad \quad \quad = -\frac{\kappa }{3}\left( \mu +3p_{r}-16c\left( \frac{\zeta }{S}\right)
^{2}\right) .  \label{nfield_zeta_1_-_nfield_zeta_3}
\end{eqnarray}
Combination of (\ref{nfield_zeta1_+_nfield_zeta_3}) with the field equations
(cf\ eq.\ (\ref{combined_field4_and_addition})) leads to 
\begin{eqnarray}
0 &\stackrel{\texttt{\rm(\ref{nfield_zeta_t_4})}}{=}&\frac{\kappa }{3}\frac{d}{dt}\left( \mu -3p_{r}+8c\left( \frac{\zeta }{S}%
\right) ^{2}\right)  \nonumber \\
&=&\frac{\kappa }{3}\frac{d}{dt}\left( \Sigma ^{\gamma }{}_{\gamma
}+8c\left( \frac{\zeta }{S}\right) ^{2}\right) \Rightarrow \Sigma ^{\gamma
}{}_{\gamma }+8c\left( \frac{\zeta }{S}\right) ^{2}=\texttt{\rm const}=:\Xi
\label{conserverd_energy_momentum_trace_plus_additional_stress}
\end{eqnarray}
Thus, we obtained a conserved quantity similar to the one in (\cite{PuetzTres}, eq.\ (47)). The first Noether identity (\ref{nnoether_zeta_t_1})
takes the form 
\begin{eqnarray}
&&\frac{d}{dt}\left[ S^{4}\left( \mu -8c\left( \frac{\zeta }{S}\right)
^{2}\right) \right] +4cS^{2}\frac{d\zeta ^{2}}{dt} =\frac{1}{4}\frac{dS^{4}%
}{dt}\Sigma ^{\gamma }{}_{\gamma }  \nonumber \\
&&\quad \quad \quad \stackrel{\texttt{\rm(\ref{conserverd_energy_momentum_trace_plus_additional_stress})}}{\Leftrightarrow} 4\frac{\dot{S}}{S}\left( \Xi -\mu -\frac{3}{4}\Sigma
^{\gamma }{}_{\gamma }\right) -\dot{\mu}\ =8c\left( \frac{\zeta }{S}%
\right) ^{2}\left( 2\frac{\dot{S}}{S}-\frac{\dot{\zeta}}{\zeta }\right) .
\label{noether2_for_zeta_simplified}
\end{eqnarray}
Before we proceed with the search for explicit solutions, we will collect the remaining field equations 
\begin{eqnarray}
&&\chi \left( \left( \frac{\dot{S}}{S}\right) ^{2}+\frac{k}{S^{2}}\right)
-\left( a_{4}+a_{6}\right) \kappa \left( \left( \frac{\ddot{S}}{S}\right)
^{2}-\left[ \left( \frac{\dot{S}}{S}\right) ^{2}+\frac{k}{S^{2}}\right]
^{2}\right)  \nonumber \\
&&\quad \quad \quad =\frac{\kappa }{3}\left( \mu -4c\left( \frac{\zeta }{S}\right) ^{2}\right)
,  \label{field_final_zeta_XI_<>_0_1} \\
&&\chi \left( \Lambda +\frac{\ddot{S}}{S}\right) +\left( a_{4}+a_{6}\right)
\kappa\left( \left( \frac{\ddot{S}}{S}\right) ^{2}-\left[ \left( \frac{%
\dot{S}}{S}\right) ^{2}+\frac{k}{S^{2}}\right] ^{2}\right)  \nonumber \\
&&\quad \quad \quad = -\kappa \left( p_{r}-4c\left( \frac{\zeta }{S}\right) ^{2}\right) ,
\label{field_final_zeta_XI_<>_0_2} \\
&&\frac{\ddot{S}}{S}+\left( \frac{\dot{S}}{S}\right) ^{2}+\frac{k}{S^{2}}
 =\Lambda ,  \label{field_final_zeta_XI_<>_0_3} \\
&&4\frac{\dot{S}}{S}\left( \Xi -\mu -\frac{3}{4}\Sigma ^{\gamma }{}_{\gamma
}\right) -\dot{\mu}\ =8c\left( \frac{\zeta }{S}\right) ^{2}\left( 2\frac{%
\dot{S}}{S}-\frac{\dot{\zeta}}{\zeta }\right) .
\label{field_final_zeta_XI_<>_0_4}
\end{eqnarray}
Note that the new constant $\Lambda $ in (\ref{field_final_zeta_XI_<>_0_2})
is defined via (\ref{field_final_zeta_XI_<>_0_3}). Comparison of
(\ref{field_final_zeta_XI_<>_0_3}) with the Friedman equation in standard
cosmology reveals that $\Lambda$ plays the same role as the usual cosmological
constant. Since we did not include this additional constant in our Lagrangian
right from the beginning $\Lambda$ might be termed {\it induced} cosmological
constant. Now let us exploit the
fact that we are allowed to set the constant $\Xi 
\stackrel{A}{=}0$, which leads to an additional constraint, i.e.\ 
\begin{equation}
\mu =3p_{r}-8c\left( \frac{\zeta }{S}\right) ^{2}.
\label{xi_=_0_energy_and_stresses_relation}
\end{equation}
Subsequently eq.\ (\ref{field_final_zeta_XI_<>_0_1}) turns into 
\begin{eqnarray}
\chi \left( \left( \frac{\dot{S}}{S}\right) ^{2}+\frac{k}{S^{2}}\right)
-\left( a_{4}+a_{6}\right) &\kappa &\left( \left( \frac{\ddot{S}}{S}\right)
^{2}-\left[ \left( \frac{\dot{S}}{S}\right) ^{2}+\frac{k}{S^{2}}\right]
^{2}\right)  \nonumber \\
\quad \quad \quad = \kappa \left( p_{r}-4c\left( \frac{\zeta }{S}\right) ^{2}\right) ,
\label{field_final_zeta_XI_=_0_1}
\end{eqnarray}
and the second Noether identity (\ref{field_final_zeta_XI_<>_0_4}) now reads 
\begin{equation}
4\frac{\dot{S}}{S}\mu +\dot{\mu}\ =8c\left( \frac{\zeta }{S}\right)
^{2}\left( \frac{\dot{S}}{S}+\frac{\dot{\zeta}}{\zeta }\right) .
\label{field_final_zeta_XI_=_0_4}
\end{equation}
Note that we collected all assumptions made up to this point in \tref{tabelle_0}.  
\begin{table}
\caption{Assumptions made up to this point.}
\label{tabelle_0}
\begin{indented}
\item[]\begin{tabular}{@{}lll}
\br
Ansatz/Assumption&Resulting quantity/equation&Equation\\\mr
$\tau_{\alpha \beta}=0$&Affects the form of the second field equation&\eref{field2_antisymmtetric_vanishing_spin_current}\\ 
$T^{\alpha}=\frac{1}{2} Q \wedge \vartheta^{\alpha}$&Affects the form of the connection&\eref{Weyl_Cartan_Konnektion}\\
$Q=\frac{\xi(t,r)}{S(t)}\vartheta^{\hat 0}$&Controls non-Riemannian features/&\eref{hypermomentum_with_inserted_strain_curvature},\eref{specific_ansatz_for_nonemtricity_trace_no_1},\eref{torsion_in_weyl_cartan_spacetime},\\
&Affects the form of the field equations&\eref{nexp_field_1}-\eref{exp_noether_2}\\
$Q=\frac{\zeta(t)}{S(t)}\vartheta^{\hat 0}$&Controls non-Riemannian
features/&\eref{hypermomentum_with_inserted_strain_curvature},\eref{specific_ansatz_for_nonemtricity_trace_no_2},\eref{torsion_in_weyl_cartan_spacetime},\\
&Simplifies field equations&\eref{nfield_zeta_t_1}-\eref{nnoether_zeta_t_2}\\
$a_4 \not = -a_6$&Affects the form of the second field equation&\eref{nexp_field_4}\\
$\Xi=\Sigma^{\alpha}{}_{\alpha}=0$&Relation between $\mu$ and $p_r$&\eref{xi_=_0_energy_and_stresses_relation}\\
$\Lambda$&Affects the form of the field equations&\eref{field_final_zeta_XI_<>_0_2}--\eref{field_final_zeta_XI_<>_0_3}\\\br
\end{tabular}
\end{indented}
\end{table}

\section{Solutions\label{special_case_xi_=_zeta_solutions_section}}

\subsection{$\Lambda \neq 0$ solutions\label{special_case_xi_=_zeta_solutions_lambda_<>_0_section}}

We are now going to solve eq.\ (\ref{field_final_zeta_XI_<>_0_3}) for
nonvanishing $\Lambda $. Note that eq.\ (\ref{field_final_zeta_XI_<>_0_3})
does not depend on the relation between the energy density and pressure and
therefore can be solved independently. This ODE, after a substitution, turns into a
Bernoulli ODE, which in turn can be transformed into a linear equation.
After this procedure we obtain two branches for the scale factor. They read
as follows: 
\begin{equation}
S=\pm \frac{1}{\sqrt{2\Lambda }}\sqrt{e^{-\sqrt{2\Lambda }\, t}\left( 2 k
    e^{\sqrt{2\Lambda }\, t}-\sqrt{2\Lambda }\varkappa_{1}e^{2\sqrt{2\Lambda
    }\, t}+\sqrt{2\Lambda }\varkappa _{2}\right) },  \label{scale_factor_lambda<>0_1}
\end{equation}
where $\varkappa_{1}$, and $\varkappa_{2}$ are constants. This solution for the scale factor is valid for all three possible choices of
$k$. Let us now proceed by fixing the equation of state. 

We will start with the most simple ansatz, i.e.\ with the introduction of an
additional constant $w$ into the equation of state, which parameterizes the
ratio of the energy density and the pressure in our model, 
\begin{equation}
w\,\,\mu (t)=p_{r}(t).  \label{parameter_equation_of_state_1_ansatz}
\end{equation}
Now let us derive the impact of (\ref{parameter_equation_of_state_1_ansatz})
on our set of field equations given by (\ref{field_final_zeta_XI_<>_0_1})--(%
\ref{field_final_zeta_XI_=_0_4}). Equation (\ref
{xi_=_0_energy_and_stresses_relation}) yields 
\begin{equation}
\mu =-\frac{8c}{1-3w}\left( \frac{\zeta }{S}\right) ^{2}.
\label{energy_dust_lambda_neq_0}
\end{equation}
The field equations are now given by 
\begin{eqnarray}
&&\chi \left( \left( \frac{\dot{S}}{S}\right) ^{2}+\frac{k}{S^{2}}\right)
-\left( a_{4}+a_{6}\right) \kappa \left( \left( \frac{\ddot{S}}{S}\right)
^{2}-\left[ \left( \frac{\dot{S}}{S}\right) ^{2}+\frac{k}{S^{2}}\right]
^{2}\right)  \nonumber \\
&&\quad \quad \quad \quad =-\frac{4c\kappa }{3}\left( \frac{\zeta }{S}\right) ^{2}\left( \frac{1+3w}{%
1-3w}\right) ,  \label{parametrized_field_lambda_ansatz_1_0} \\
&&\chi \left( \Lambda +\frac{\ddot{S}}{S}\right) +\left( a_{4}+a_{6}\right)
\kappa\left( \left( \frac{\ddot{S}}{S}\right) ^{2}-\left[ \left( \frac{%
\dot{S}}{S}\right) ^{2}+\frac{k}{S^{2}}\right] ^{2}\right)  \nonumber \\
&&\quad \quad \quad \quad = 4c\kappa \left( \frac{\zeta }{S}\right) ^{2}\left( \frac{1-w}{1-3w}%
\right) ,  \label{parametrized_field_lambda_ansatz_1_1} \\
&&\frac{\ddot{S}}{S}+\left( \frac{\dot{S}}{S}\right) ^{2}+\frac{k}{S^{2}}
=\Lambda ,  \label{parametrized_field_lambda_ansatz_1_2} \\
&&\frac{24c\left( 1-w\right) }{3w-1}\left( \frac{\zeta }{S}\right) ^{2}\left( 
\frac{\dot{S}}{S}+\frac{\dot{\zeta}}{\zeta }\right) =0.
\label{parametrized_field_lambda_ansatz_1_4}
\end{eqnarray}
Equation (\ref{parametrized_field_lambda_ansatz_1_4}) has two non-trivial
solution, namely 
\begin{equation}
\zeta =\frac{\iota }{S},\texttt{\rm\thinspace \thinspace with }\iota
=\texttt{\rm const,}\qquad \texttt{\rm and \ } \quad w=1.  \label{parametrized_first_condition}
\end{equation}
Solving the remaining field equations with respect to the first solution in
eq.\ (\ref{parametrized_first_condition}), we obtain constraints
among the coupling constants which are summarized in \tref{tabelle_1} (note that every
set of parameters on the rhs corresponds to a solution of the field equations).
\begin{table}
\caption{Ansatz $\Lambda \neq 0, w=\texttt{\rm const}$.}
\label{tabelle_1}
\begin{indented}
\item[]\begin{tabular}{@{}lll}
\br
$\zeta$&$S$&Additional constraints\\\mr
$\zeta=\frac{\iota}{S}$&$S$ from eq.\
(\ref{scale_factor_lambda<>0_1})&$\{a_{4}=a_{6}=c=\chi =0\},$\\
&& $\{a_{4}=a_{6}=\iota =0,w=\frac{1}{3}\},$\\ 
&&$\{a_{4}=a_{6}=\chi =0,w=1\},$\\ 
&&$\{a_{4}=a_{6}=c=0,$ $w=\frac{1}{3}\}$,\\
&& $\{a_{4}=-a_{6},\Lambda =0, w= \texttt{\rm const}\}$  \\ \br
\end{tabular}
\end{indented}
\end{table}
These solutions are not very satisfactory since they either lead to
vanishing post-Riemannian quantities or, in case of $\chi =c=0$, to a
restriction on the Lagrangian level. 

Let us switch to another ansatz for the equation of state, namely 
\begin{equation}
w(t)\mu (t)=p_{r}(t).  \label{parameter_equation_of_state_2_ansatz}
\end{equation}
Thus, we introduced an additional function into the equation of state which
controls the relation between the energy density and stresses in a \textit{%
dynamical} way. The field equation which changes with respect to the set (%
\ref{parametrized_field_lambda_ansatz_1_0})--(\ref
{parametrized_field_lambda_ansatz_1_4}), besides of the fact that $w$ is no
longer a constant, is the Noether identity in eq.\ (\ref
{parametrized_field_lambda_ansatz_1_4}), which now reads 
\begin{eqnarray}
\hspace{-1cm}&&-\frac{24c}{\left( 3w-1\right) ^{2}}\left( \frac{\zeta }{S%
}\right) ^{2}\left( \frac{\dot{S}}{S}\left( 1+3w^{2}-4w\right) +\frac{\dot{%
\zeta}}{\zeta }\left( 1+3w^{2}-4w\right) +\dot{w}\right) =0.
\label{parametrized_field_lambda_ansatz_2_4}
\end{eqnarray}
In case of an arbitrary choice of $\zeta$, this equation is solved by 
\begin{equation}
w=\frac{S^{2}\zeta ^{2}\varkappa _{3}-1}{S^{2}\zeta ^{2}\varkappa _{3}-3}.
\label{solution_for_w_lambda<>0_case_ansatz_2}
\end{equation}
Reinsertion of this solution for $w$ into the remaining field equations yields
additional parameter constraints which are summarized in \tref{tabelle_2}. 
As one realizes immediately, none of the solutions collected in \tref{tabelle_2} is of use for
us, since they all lead to unrealistic or forbidden restrictions among the
coupling constants in our model. Therefore, in the following section, we will
switch to the case in which the induced cosmological constant $\Lambda$ vanishes.  
\begin{table}
\caption{Ansatz $\Lambda \neq 0,w=w(t)$.}
\label{tabelle_2}
\begin{indented}
\item[]\begin{tabular}{@{}llll}
\br
$\zeta $ & $S$ & Additional constraints &  \\\mr
$\zeta $ arbitrary, $w=\frac{\zeta ^{2}S^{2}\varkappa _{3}-1}{\zeta ^{2}S^{2}\varkappa _{3}-3}$ & $S$ from eq.\ (\ref{scale_factor_lambda<>0_1}) & $%
\{a_{4}=-a_{6},\Lambda =0\},$ &  \\ 
&  & $\{a_{4}=-a_{6},\chi =0,w=1\},$ &  \\ 
&  & $\{a_{4}=-a_{6},\iota =\chi =0\}$ & \\\br
\end{tabular}
\end{indented}
\end{table}

\subsection{$\Lambda =0$ solutions\label{special_case_xi_=_zeta_solutions_lambda_=_0_section}}
Solving equation (\ref{field_final_zeta_XI_<>_0_3}) for vanishing $\Lambda$, 
yields a solution for the scale factor which depends on the value of the constant $k$: 
\begin{eqnarray}
k &\not=&0:\quad S=\pm \sqrt{\frac{1}{k}\left( \varkappa _{1}-k^{2}\left(
\varkappa _{2}+t\right) ^{2}\right) }\, ,\quad \texttt{\rm with \quad }\varkappa
_{1},\varkappa _{2}=\texttt{\rm const,}
\label{special_vacuum_solution_scale_factor_k<>0} \\
k &=&0:\quad S=\varkappa _{1}\texttt{\rm or }S=\pm \,\sqrt{2\varkappa _{1}\left(
t+\varkappa _{2}\right) }\, ,\quad \texttt{\rm with\quad }\varkappa _{1},\varkappa
_{2}=\texttt{\rm const}.
\label{special_vacuum_solution_scale_factor_k=0}
\end{eqnarray}
Motivated by the results in the previous section for the $\Lambda \not=0$
case, we will directly start with the more general
equation of state as given in (\ref{parameter_equation_of_state_2_ansatz}). The field
equations are now given by eqs.\ (\ref{parametrized_field_lambda_ansatz_1_0})-(
\ref{parametrized_field_lambda_ansatz_1_4}) but with $\Lambda =0$. The
parameter constraints for this solution are summarized in the second part of \tref{tabelle_3}.

Additionally, we investigated the case in which made use of the old solution
for $\zeta $, i.e.\ as given in (\ref{parametrized_first_condition}). With
this ansatz for $\zeta $ the second Noether identity, as given in eq.\ (\ref
{parametrized_field_lambda_ansatz_2_4}), turns into: 
\begin{equation}
\dot{w}=0.  \label{second_noether_for_old_zeta_=_iota_div_S_ansatz}
\end{equation}
Thus, $w$ has to be a constant which, subsequently, can be determined from
the remaining field equations after choosing the branch for $S$ from eqs.\ (%
\ref{special_vacuum_solution_scale_factor_k<>0})--(\ref
{special_vacuum_solution_scale_factor_k=0}). The additional constraints for
this parameter choice are listed in the first part of \tref{tabelle_3}. Most
interestingly it turns out that the parameter $w$, which controls the
equation of state, is restricted by the choice of a certain set of constants
in our theory (cf rhs in \tref{tabelle_3}).    
\begin{table}
\caption{Ansatz $\Lambda = 0$, $w=w(t)$.}
\label{tabelle_3}
\begin{indented}
\item[]\begin{tabular}{@{}llll}
\br
$\zeta $ & $S$ & Additional constraints &  \\ \mr
$\zeta =\frac{\iota }{S},$ $w=$ const & $k\not=0,$ $S$ from eq.\ (\ref
{special_vacuum_solution_scale_factor_k<>0}) & $w=\frac{4c\iota ^{2}\kappa
+\varkappa _{1}\chi }{4c\iota ^{2}\kappa +3\varkappa _{1}\chi }$ &  \\ 
& $k=0,$ $S=$ const cf\ eq.\ (\ref{special_vacuum_solution_scale_factor_k=0})
& $w=1$ &  \\ 
& $k=0,$ $S$ from eq.\ (\ref{special_vacuum_solution_scale_factor_k=0}) & $w=%
\frac{4c\iota ^{2}\kappa +\varkappa _{1}^{2}\chi }{4c\iota ^{2}\kappa
+3\varkappa _{1}^{2}\chi }$ &  \\ \mr
$w=\frac{\zeta ^{2}S^{2}\varkappa _{3}-1}{\zeta ^{2}S^{2}\varkappa _{3}-3}$, 
$\zeta $ arbitrary & $k\not=0,$ $S$ from eq.\ (\ref
{special_vacuum_solution_scale_factor_k<>0}) & $\varkappa _{3}=-4\frac{%
c\kappa }{\varkappa _{1}\chi }$ &  \\ 
& $k=0,$ $S=$ const cf\ eq.\ (\ref{special_vacuum_solution_scale_factor_k=0})
& $c=0$ &  \\ 
& $k=0,$ $S$ from eq.\ (\ref{special_vacuum_solution_scale_factor_k=0}) & $%
\varkappa _{3}=-4\frac{c\kappa }{\varkappa _{1}^{2}\chi }$ & \\ \br
\end{tabular}
\end{indented}
\end{table}

\begin{figure}[ht]
\setlength{\unitlength}{1mm}
\begin{picture}(80,50)
\epsfig{file=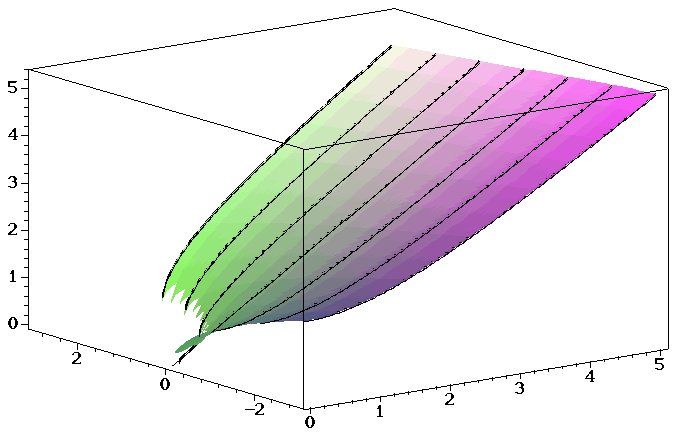}
\put(-60,4){\tiny{$\kappa_1$}}
\put(-70,30){\tiny{$\kappa_2=0$, $k=-1$}} 
\put(-77,26){\tiny{$S$}}
\put(-25,4){\tiny{$t$}}
\end{picture}
\begin{picture}(80,50)
\epsfig{file=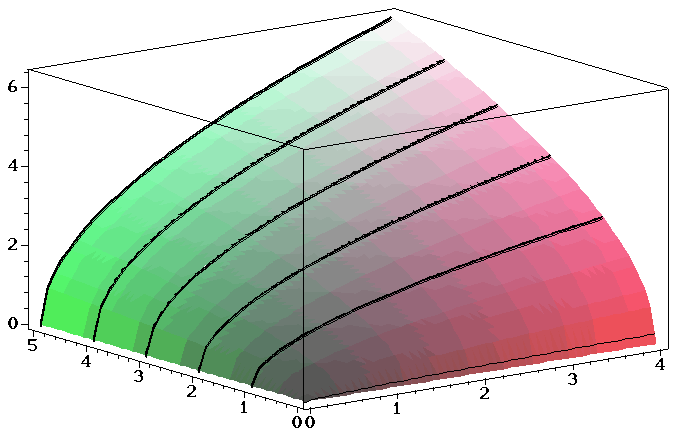}
\put(-60,4){\tiny{$\kappa_1$}}
\put(-72,33){\tiny{$\kappa_2 = k = 0$}} 
\put(-77,26){\tiny{$S$}}
\put(-25,4){\tiny{$t$}}
\end{picture}
\begin{picture}(80,50)
\epsfig{file=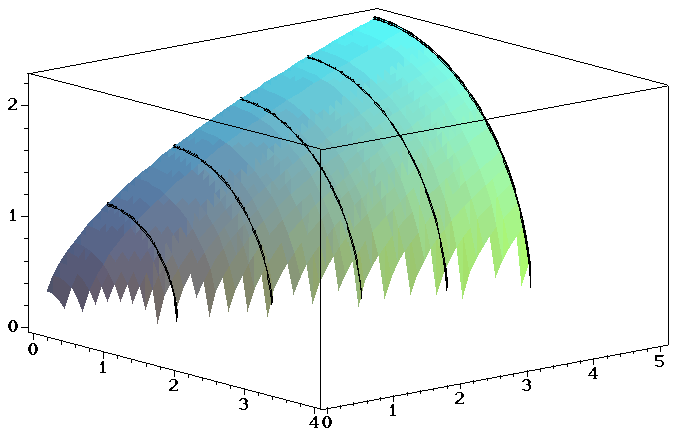}
\put(-25,4){\tiny{$\kappa_1$}}
\put(-34,12){\tiny{$\kappa_2=0$, $k=1$}} 
\put(-77,26){\tiny{$S$}}
\put(-60,4){\tiny{$t$}}
\end{picture}
\begin{picture}(80,50)
\epsfig{file=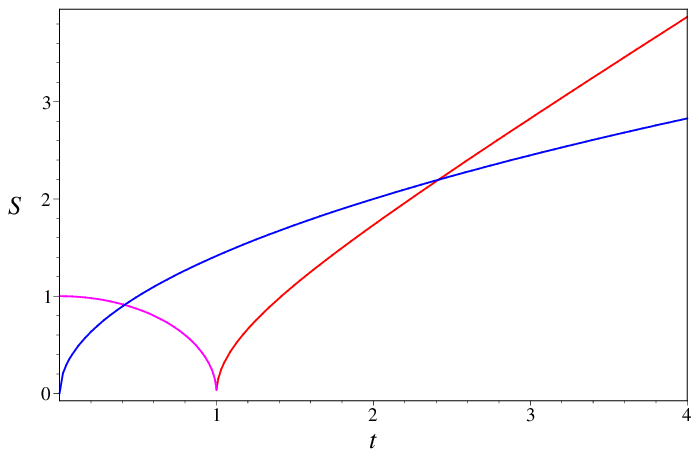}
\put(-16,26){\tiny{$k=0$}}
\put(-20,38){\tiny{$k=-1$}}
\put(-58,8){\tiny{$k=1$}}
\put(-63,40){\tiny{$\kappa_1=\kappa_2=1$}} 
\end{picture}
\caption[Scale factors for $\Lambda=0$ solutions]{Temporal
  behaviour of the scale factor in the case of the
  $\Lambda=0$ branch of the model (we always select the positive sign in front
  of the scale factor, cf\ eqs.\ (\ref{special_vacuum_solution_scale_factor_k<>0})--(\ref{special_vacuum_solution_scale_factor_k=0})).}\label{fig_scale_factors_Lambda_0}
\end{figure}

\begin{figure}[ht]
\setlength{\unitlength}{1mm}
\begin{picture}(80,50)
\epsfig{file=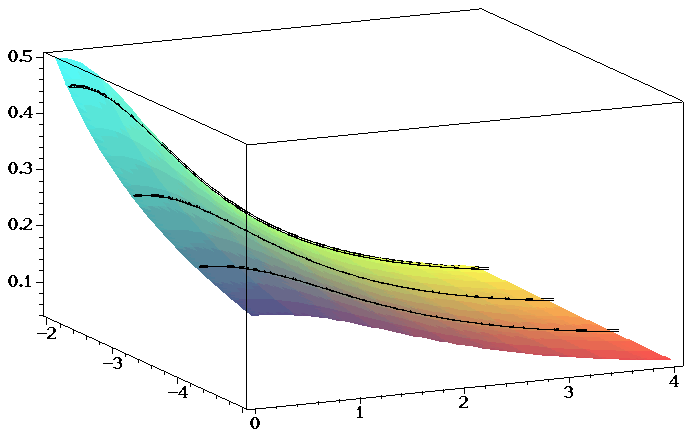}
\put(-34,29){\tiny{$\kappa_2=0$, $k=-1$, $\iota=1$}} 
\put(-64,7){\tiny{$\kappa_1$}}
\put(-30,3){\tiny{$t$}}
\put(-79,30){\tiny{$Q$}}
\end{picture}
\begin{picture}(80,50)
\epsfig{file=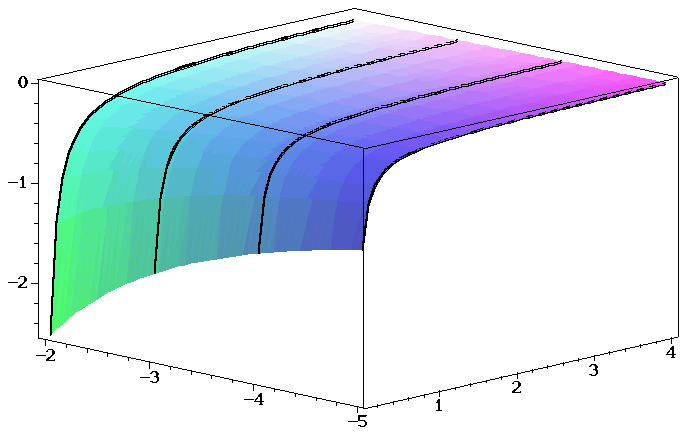}
\put(-30,16){\tiny{$\kappa_2 = k = 0$, $\iota=1$}} 
\put(-59,6){\tiny{$\kappa_1$}}
\put(-20,6){\tiny{$t$}}
\put(-77,25){\tiny{$Q$}}
\end{picture}
\begin{picture}(80,50)
\epsfig{file=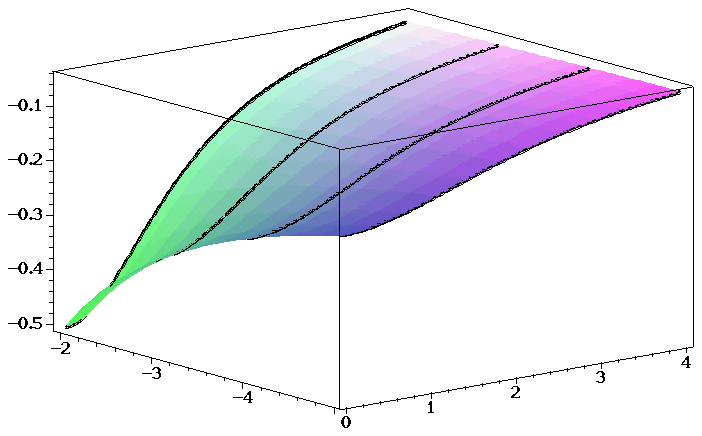}
\put(-34,14){\tiny{$\kappa_2=0$, $k=1$, $\iota=1$}} 
\put(-60,6){\tiny{$\kappa_1$}}
\put(-22,6){\tiny{$t$}}
\put(-79,27){\tiny{$Q$}}
\end{picture}
\begin{picture}(80,50)
\epsfig{file=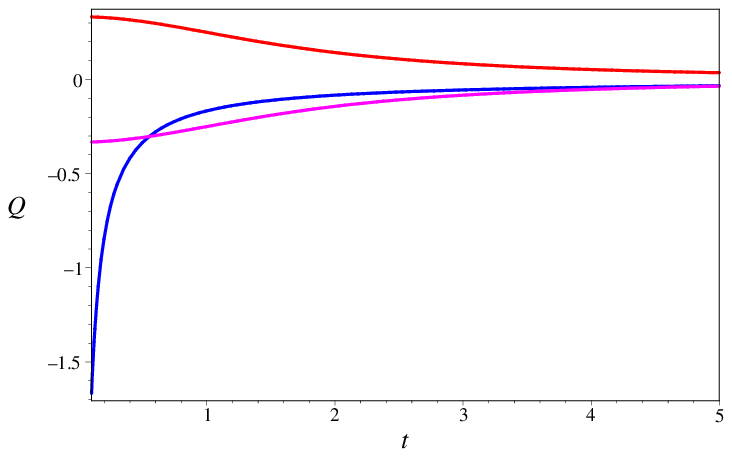}
\put(-35,31){\tiny{$k=1$}}
\put(-19,41){\tiny{$k=-1$}}
\put(-62,12){\tiny{$k=0$}}
\put(-30,10){\tiny{$\kappa_1=-3$, $=\kappa_2=0$, $\iota=1$}} 
\end{picture}
\caption[Weyl 1-form $Q$ for $\Lambda=0$ solutions]{Temporal behaviour of the
  non-Riemannian Weyl 1-form $Q$ in the case of the
  $\Lambda=0$, $\zeta=\frac{\iota}{S}$ branch of the model (we always select the positive sign in front
  of the scale factor, cf\ eqs.\
  (\ref{special_vacuum_solution_scale_factor_k<>0})--(\ref{special_vacuum_solution_scale_factor_k=0})).}\label{fig_Q_factors_Lambda_0}
\end{figure}

\section{Conclusion \label{CONCLUSION_SECTION}}

In the the last section we have shown that it is possible to find exact solutions of
the field equations within our model. We were able to generate a rather broad class of
solutions which allows for a flexible equation of state. We collected the
resulting constraints of the parameters in our model in tables
\ref{tabelle_1}--\ref{tabelle_3}. There seem to be {\it no} reasonable
solutions in case of a non-vanishing induced cosmological constant
$\Lambda$, unless one wants to introduce strong restrictions on the Lagrangian level. 
Thus, we are going to focus on the solutions with {\em vanishing} $\Lambda$ in the following. 

In \fref{fig_scale_factors_Lambda_0} we plotted the scale factor for all three
possible values of $k$ and for different values of the parameter $\varkappa_1$.
As becomes clear from the plot at bottom right, we have three qualitatively
different behaviours depending on the value of $k$. As in the Friedman case
the collapsing scenario corresponds to a universe with positive spatial
curvature. In \fref{fig_Q_factors_Lambda_0} we plotted the function
$Q$ for the ansatz mentioned in equation \eref{parametrized_first_condition}.
As stated before $Q$, the Weyl 1-form controls the non-Riemannian features of our model. From
the plots it becomes clear that it is possible to construct models in which $Q$
vanishes at later times. Thus, the non-Riemmanian quantities {\it die out} with
time. This is a rather desirable behaviour, since the spacetime we are living in
nowadays seems to be a Riemannian one. At least all experiments 
carried out so far point into this direction \cite{ProceedingsHonnef}. 
Nevertheless our model is flexible enough to cope with both situations, i.e.\ if there
is evidence for non-Riemannian structures at the present time, we are able to implement 
this fact by modifying our ansatz in \eref{specific_ansatz_for_nonemtricity_trace_no_1}
and \eref{specific_ansatz_for_nonemtricity_trace_no_2}, respectively.

In comparison with the usual FLRW model of cosmology we still have three
distinct cases for the evolution of the scale factor, which correspond to
the three different choices for $k$ in the ansatz for the metric in
equation (\ref{robertson_walker_coframe}). Since one of our field equations
(\ref{field_final_zeta_XI_<>_0_3}) is very similar to the
Friedman equation in standard cosmology we obtain a similar root type
behaviour for the scale factor as displayed in figure \ref{fig_scale_factors_Lambda_0}. As shown in (\ref{scale_factor_lambda<>0_1}) an
induced cosmological constant leads to inflationary like solutions. In
contrast to our old model \cite{PuetzTres} we were not able to find meaningful
parameter constraints for this branch of the model (cf. tables \ref{tabelle_1}
and \ref{tabelle_2}). This
drawback might be relaxed in the future if we switch to another ansatz for the
Weyl 1-form $Q$. Most interestingly the non-Riemannian quantities lead to a
contribution to the total energy density of the universe as shown in
(\ref{energy_dust_lambda_neq_0}). Thus, the energy density $\mu$ is no longer
a quantity which is determined by the evolution of the scale factor only, like in the FLRW scenario. As
we will show in the next article of this series this contribution might be
used to define a new energy density parameter which adds to the total energy
budget of the universe. Thereby leading to an interesting new source for a
possible dark energy component. Since the field equations differ from the
Friedman equations one can expect several observational changes with respect
to the standard FLRW model. Note that an ansatz with a position dependent Weyl
1-form at very early stages of the universe might contribute to the observed
inhomogeneities in the cosmic microwave background. Although speculative at
this time, small inhomogeneities in the new geometric quantities might also
have served as seeds for structure formation at early stages, thereby yielding an
interesting supplement to the quantities within the standard paradigm.     

Let us summarize that the solutions found above contribute to the collection
of known exact solutions in MAG, see \cite{Exact2}. Additionally, we managed
to extend the model proposed in \cite{PuetzTres}. We provided the foundation
for upcoming articles which will deal with the observational consequences of
this new model. The most pressing task will be to look for realistic parameter choices in order to determine whether the model is in agreement with recent
observational data. Within the next article of this series \cite{ExtweylObs},
we will use the supernova data of Perlmutter \etal \cite{Perlmutter} and
Schmidt \etal \cite{Schmidt} in order to constrain the free parameters in our model.

\appendix

\section{MAG in general\label{MAG_KAPITEL}}

In MAG we have the metric $g_{\alpha \beta }$, the coframe $\vartheta
^{\alpha }$, and the connection 1-form $\Gamma _{\alpha}{}^{\beta}$ (with
values in the Lie algebra of the four-dimensional linear group $GL(4,R)$) as
new independent field variables.\ Here $\alpha ,\beta ,\ldots =0,1,2,3$
denote (anholonomic) frame indices. Spacetime is described by a
metric-affine geometry with the gravitational field strengths nonmetricity $%
Q_{\alpha \beta }:=-Dg_{\alpha \beta }$, torsion $T^{\alpha }:=D\vartheta
^{\alpha }$, and curvature $R_{\alpha }{}^{\beta }:=d\Gamma _{\alpha
}{}^{\beta }-\Gamma _{\alpha }\,^{\gamma }\wedge \Gamma _{\gamma }{}^{\beta
} $. A Lagrangian formalism for a matter field $\Psi $ minimally coupled to
the gravitational potentials $g_{\alpha \beta }$, $\vartheta ^{\alpha }$, $%
\Gamma _{\alpha }{}^{\beta }$ has been set up in \cite{PhysRep}. 
The dynamics of an ordinary MAG theory is specified by a total Lagrangian 
\begin{equation}
L=V_{{\rm MAG}}(g_{\alpha \beta },\vartheta ^{\alpha },Q_{\alpha \beta
},T^{\alpha },R_{\alpha }{}^{\beta })+L_{{\rm mat}}(g_{\alpha \beta
},\vartheta ^{\alpha },\Psi ,D\Psi ).
\end{equation}
The variation of the action with respect to the independent gauge potentials
leads to the field equations: 
\begin{eqnarray}
\frac{\delta L_{{\rm mat}}}{\delta \Psi } &=&0,  \label{matter} \\
DM^{\alpha \beta }-m^{\alpha \beta } &=&\sigma ^{\alpha \beta },
\label{zeroth} \\
DH_{\alpha }-E_{\alpha } &=&\Sigma _{\alpha ,}  \label{first} \\
DH^{\alpha }{}_{\beta }-E^{\alpha }{}_{\beta } &=&\Delta ^{\alpha }{}_{\beta
}.  \label{second}
\end{eqnarray}
Equations (\ref{zeroth}) and (\ref{first}) are the generalized Einstein
equations with the symmetric
energy-momentum 4-form $\sigma^{\alpha \beta}$ and the canonical
energy-momentum 3-form $\Sigma _{\alpha }$ as sources. Equation (\ref{second})
is an additional field equation which takes into
account other aspects of matter, such as spin, shear and dilation currents,
represented by the hypermomentum $\Delta ^{\alpha }{}_{\beta }$. We made use
of the definitions of the gauge field excitations, 
\begin{equation}
H_{\alpha }:=-\frac{\partial V_{{\rm MAG}}}{\partial T^{\alpha }},\quad
H^{\alpha }{}_{\beta }:=-\frac{\partial V_{{\rm MAG}}}{\partial R_{\alpha
}{}^{\beta }},\quad M^{\alpha \beta }:=-2\frac{\partial V_{{\rm MAG}}}{%
\partial Q_{\alpha \beta }},  \label{exications}
\end{equation}
of the canonical energy-momentum, the metric stress-energy, and the
hypermomentum current of the gauge fields, 
\begin{equation}
E_{\alpha }:=\frac{\partial V_{{\rm MAG}}}{\partial \vartheta ^{\alpha }}%
,\quad m^{\alpha \beta }:=2\frac{\partial V_{{\rm MAG}}}{\partial g_{\alpha
\beta }},\quad E^{\alpha }{}_{\beta }=-\vartheta ^{\alpha }\wedge H_{\beta
}-g_{\beta \gamma }M^{\alpha \gamma },  \label{gauge_currents}
\end{equation}
and of the canonical energy-momentum, the metric stress-energy, and the
hypermomentum currents of the matter fields, 
\begin{equation}
\Sigma _{\alpha }:=\frac{\delta L_{{\rm mat}}}{\delta \vartheta ^{\alpha }}%
,\quad \sigma ^{\alpha \beta }:=2\frac{\delta L_{{\rm mat}}}{\delta
g_{\alpha \beta }},\quad \Delta ^{\alpha }{}_{\beta }:=\frac{\delta L_{{\rm %
mat}}}{\delta \Gamma _{\alpha }{}^{\beta }}.  \label{matter_currents}
\end{equation}
Provided the matter equations (\ref{matter}) are fulfilled, the following
Noether identities hold: 
\begin{eqnarray}
D\Sigma _{\alpha } &=&\left( e_{\alpha }\rfloor T^{\beta }\right) \wedge
\Sigma _{\beta }-\frac{1}{2}\left( e_{\alpha }\rfloor Q_{\beta \gamma
}\right) \sigma ^{\beta \gamma }+\left( e_{\alpha }\rfloor R_{\beta
}{}^{\gamma }\right) \wedge \Delta ^{\beta }{}_{\gamma },
\label{noether_ident_1} \\
D\Delta ^{\alpha }{}_{\beta } &=&g_{\beta \gamma }\sigma ^{\alpha \gamma
}-\vartheta ^{\alpha }\wedge \Sigma _{\beta }.  \label{noether_ident_2}
\end{eqnarray}
They show that the field equation (\ref{zeroth}) is redundant, thus we only
need to take into account (\ref{first}) and (\ref{second}).

As suggested in \cite{Exact2}, the most general parity conserving quadratic
Lagrangian expressed in terms of the irreducible pieces of the nonmetricity $%
Q_{\alpha \beta }$, torsion $T^{\alpha }$, and curvature $R_{\alpha \beta }$
reads 
\begin{eqnarray}
V_{{\rm MAG}}=&\frac{1}{2\kappa }\biggl[ &-a_{0}R^{\alpha \beta }\wedge \eta
_{\alpha \beta }-2\lambda \eta +T^{\alpha }\wedge \,^{\star }\left(
\sum_{I=1}^{3}a_{I}\,^{(I)}T_{\alpha }\right)   \nonumber \\
&&+Q_{\alpha \beta }\wedge \,^{\star }\left(
\sum_{I=1}^{4}b_{I}\,^{(I)}Q^{\alpha \beta }\right)\nonumber \\  &&+b_{5}\left(
^{(3)}Q_{\alpha \gamma }\wedge \vartheta ^{\alpha }\right) \wedge \,^{\star
}\left( \,^{(4)}Q^{\beta \gamma }\wedge \vartheta _{\beta }\,\right)   \nonumber
\\
&&+2\left( \sum_{I=2}^{4}c_{I}\,^{(I)}Q_{\alpha \beta }\right) \wedge
\vartheta ^{\alpha }\wedge \,^{\star }T^{\beta }\biggr]  \nonumber \\
-\frac{1}{2\rho }\, R^{\alpha \beta }\wedge \,^{\star
}&&\hspace{-0.5cm}\biggl[\sum_{I=1}^{6}w_{I}\,^{(I)}W_{\alpha \beta
}+\sum_{I=1}^{5}z_{I}\,^{(I)}Z_{\alpha \beta }+w_{7}\vartheta _{\alpha
}\wedge \left( e_{\gamma }\rfloor \,^{(5)}W^{\gamma }{}_{\beta }\right)  
\nonumber \\
&&\hspace{-0.5cm}+z_{6}\vartheta _{\gamma }\wedge \left( e_{\alpha }\rfloor
\,^{(2)}Z^{\gamma }{}_{\beta }\right) +\sum_{I=7}^{9}z_{I}\vartheta _{\alpha
}\wedge \left( e_{\gamma }\rfloor \,^{(I-4)}Z^{\gamma }{}_{\beta }\right) \biggr].
\label{general_v_mag}
\end{eqnarray}
The constants entering (\ref{general_v_mag}) are the cosmological constant $%
\lambda $, the weak and strong coupling constant $\kappa $ and $\rho $\footnote{$[\lambda ]=$length$^{-2}$, $[\kappa ]=$length$^{2}$, $[\rho]=[\hbar ]=[c]=1.$}, and the 28 dimensionless parameters 
\begin{equation}
a_{0},\dots ,a_{3},b_{1},\dots ,b_{5},c_{2},\dots ,c_{4},w_{1},\dots
,w_{7},z_{1},\dots ,z_{9}.  \label{general_coupling}
\end{equation}
This Lagrangian and the presently known exact solutions in MAG have been
reviewed in \cite{Exact2}. We note that this Lagrangian incorporates the one
used in section \ref{TRESGUERRES_KAPITEL} eq.\ (\ref{Langrangian_new_symbolic}), as can be seen easily by making the following choice for the constants in
(\ref{general_v_mag}): 
\begin{equation}
\lambda ,a_{1},\dots ,a_{3},b_{1},\dots ,b_{3},b_{5},c_{2},\dots,c_{4},w_{7},z_{1},\dots ,z_{3},z_{5},\dots ,z_{9}=0.\quad 
\end{equation}
In order to obtain exactly the form of (\ref{Langrangian_new_symbolic}), one
has to perform the additional substitutions: 
\begin{equation}
a_{0}\rightarrow -\chi ,\, w_{1},\dots ,w_{6}\rightarrow -2\rho
a_{1},\dots ,-2\rho a_{6},\, b_{4}\rightarrow c,\, z_{4}\rightarrow -2\rho b.
\end{equation}
In table \ref{tabelle_9} we collected some of symbols
defined within this appendix.
\begin{table}
\caption{Summary of definitions made in \ref{MAG_KAPITEL}.}
\label{tabelle_9}
\begin{indented}
\item[]\begin{tabular}{@{}lllll}
\br
Potentials&Field strengths&Excitations&Gauge currents&\\ 
\mr
{$g_{\alpha\beta}$}&{$Q_{\alpha \beta }:=-Dg_{\alpha \beta}$}& $M^{\alpha
  \beta}:=-2\frac{\partial V}{\partial Q_{\alpha \beta}}$
&$m^{\alpha \beta}:=2\frac{\partial V}{\partial g_{\alpha \beta}}$  & \\
{$\vartheta^\alpha$}&{$T^{\alpha }\hspace{0.2cm}:=D\vartheta^\alpha$}&
$H_\alpha\hspace{0.22cm}:=-\frac{\partial V}{\partial T^{\alpha}}$ &
$E_\alpha\hspace{0.21cm}:=\frac{\partial V}{\partial \vartheta^{\alpha}}$ & \\
{$\Gamma_{\alpha}{}^{\beta}$}&{$R_{\alpha}{}^{\beta
    }:=$''$D$''$\Gamma_{\alpha}{}^{\beta}$}&
$H^{\alpha}{}_{\beta}:=-\frac{\partial V}{\partial R_{\alpha}{}^{\beta}}$ &
$E^{\alpha}{}_{\beta}:=\frac{\partial V}{\partial \Gamma_{\alpha}{}^{\beta}}$ &\\
\br
\end{tabular}
\end{indented}
\end{table}

\section{Weyl-Cartan spacetime\label{WEYL_CARTAN_KAPITEL}}

The Weyl-Cartan spacetime ($Y_{n}$) is a special case of the general
metric-affine geometry in which the tracefree part $Q_{\alpha \beta
  }\!\!\!\!\!\!\!\!\!\!\!\!\!\nearrow \,\,\,\,\,\,\,$ of the nonmetricity
$Q_{\alpha \beta }\,$ vanishes. Thus, the whole
nonmetricity is proportional to its trace part, i.e. the Weyl 1-form $Q:=%
\frac{1}{4}Q^{\alpha }{}_{\alpha }$, 
\begin{equation}
Q_{\alpha \beta }=g_{\alpha \beta }\,Q=\frac{1}{4}g_{\alpha \beta
}\,Q^{\gamma }{}_{\gamma }.  \label{torsion_in_weyl_cartan_spacetime}
\end{equation}
Therefore the general MAG connection reduces to 
\begin{eqnarray}
\Gamma _{\alpha \beta } &=&\frac{1}{2}dg_{\alpha \beta }+\left( e_{[\alpha
}\rfloor dg_{\beta ]\gamma }\right) \vartheta ^{\gamma }+e_{[\alpha }\rfloor
C_{\beta ]}-\frac{1}{2}\left( e_{\alpha }\rfloor e_{\beta }\rfloor C_{\gamma
}\right) \vartheta ^{\gamma }  \nonumber \\
&&-e_{[\alpha }\rfloor T_{\beta ]}+\frac{1}{2}\left( e_{\alpha }\rfloor
e_{\beta }\rfloor T_{\gamma }\right) \vartheta ^{\gamma }+\frac{1}{2}%
g_{\alpha \beta }\,Q+\left( e_{[\alpha }\rfloor Q\right) \vartheta _{\beta ]}
\\
&=&\Gamma _{\alpha \beta }^{\{\,\}}-e_{[\alpha }\rfloor T_{\beta ]}+\frac{1}{%
2}\left( e_{\alpha }\rfloor e_{\beta }\rfloor T_{\gamma }\right) \vartheta
^{\gamma }+\frac{1}{2}g_{\alpha \beta }\,Q+\left( e_{[\alpha }\rfloor
Q\right) \vartheta _{\beta ]}.  \label{Weyl_Cartan_Konnektion}
\end{eqnarray}
Thus, it does not include any more a symmetric tracefree part. Let us now
recall the definition of the material hypermomentum $\Delta _{\alpha \beta }$
given in (\ref{matter_currents}). Due to the absence of a symmetric
tracefree piece in (\ref{Weyl_Cartan_Konnektion}), $\Delta _{\alpha \beta }$
decomposes as follows 
\begin{eqnarray}
\Delta _{\alpha \beta } &=&\texttt{\rm antisymmetric piece + trace piece}  \nonumber
\\
&=&\tau _{\alpha \beta }+\frac{1}{4}g_{\alpha \beta }\,\Delta =\tau _{\alpha
\beta }+\frac{1}{4}g_{\alpha \beta }\,\Delta ^{\gamma }{}_{\gamma }  \nonumber
\\
&=&\texttt{\rm spin current + dilation current.}  \label{Weyl_Cartan_hypermomentum}
\end{eqnarray}
According to (\ref{Weyl_Cartan_hypermomentum}) the second MAG field equation
(\ref{second}) decomposes into 
\begin{eqnarray}
dH^{\alpha }{}_{\alpha }-E^{\alpha }{}_{\alpha } &=&\Delta , \\
g_{\gamma \lbrack \alpha }DH^{\gamma }\,_{\beta ]}-E_{[\alpha \beta ]}
&=&\tau _{\alpha \beta },
\end{eqnarray}
while the first field equation is still given by (\ref{first}).
Additionally, we can decompose the second Noether identity (\ref
{noether_ident_2}) into 
\begin{eqnarray}
\frac{1}{4}g_{\alpha \beta }\,d\Delta +\vartheta _{(\alpha }\wedge \Sigma
_{\beta )} &=&\sigma _{\alpha \beta },  \label{weyl_cartan_second_noether_1}
\\
D\tau _{\alpha \beta }+Q\wedge \tau _{\alpha \beta }+\vartheta _{\lbrack
\alpha }\wedge \Sigma _{\beta ]} &=&0.  \label{weyl_cartan_second_noether_2}
\end{eqnarray}
Thus, the first Noether identity (\ref{noether_ident_1}) with inserted Weyl
1-form and hypermomentum reads 
\begin{equation}
D\Sigma _{\alpha }=\left( e_{\alpha }\rfloor T^{\beta }\right) \wedge \Sigma
_{\beta }-\frac{1}{2}\left( e_{\alpha }\rfloor Q\right) \sigma ^{\beta
}{}_{\beta }+\left( e_{\alpha }\rfloor R_{[\beta \gamma ]}{}\right) \wedge
\tau ^{\beta \gamma }+\frac{1}{4}\left( e_{\alpha }\rfloor R\right) \wedge
\Delta .  \label{connection_weyl_cartan_final}
\end{equation}
Finally, we note that in a $Y_{n}$ spacetime the symmetric part of the
curvature $R_{(\alpha \beta )}=Z_{\alpha \beta }\,$, i.e. the strain
curvature, reduces to the trace part 
\begin{equation}
Z_{\alpha \beta }=\frac{1}{4}g_{\alpha \beta }R=\frac{1}{4}g_{\alpha \beta
}R^{\gamma }{}_{\gamma }=\frac{1}{2}g_{\alpha \beta }\,dQ. \label{strain_curvature_in_weyl_cartan_general}
\end{equation}

\section{Differential geometric formalism\label{OPERATIONS_SECTION}}
We assume a connected $n$-dimensional differential manifold $Y_n$ as
underlying structure throughout the paper. A vector basis of its tangent space
$T_p Y_n$ is denoted by $e_\alpha$, which is dual (i.e. $e_\alpha \rfloor
\vartheta^\beta=\delta_\alpha^\beta$) to the basis
$\vartheta^{\alpha}$ of the cotangent space $T_p^{\ast} Y_n$. A $p$-form
$\Xi$ can be expanded with respect to this basis as follows
\begin{equation}
\Xi=\frac{1}{p!}\, \Xi_{\beta_1 \dots \beta_p} \, \vartheta^{\beta_1} \wedge
\dots \wedge \vartheta^{\beta_p}.
\end{equation}
Table \ref{tabelle_8} provides a rough overview of the operators used
throughout the paper. For a more comprehensive treatment the reader should
consult \cite{Nakahara} or section 3, and Appendix A of \cite{PhysRep}.   
\begin{table}
\caption{Operators.}
\label{tabelle_8}
\begin{indented}
\item[]\begin{tabular}{@{}lllll}
\br
Operation&Symbol&Input&&Output\\ 
\mr
Exterior multiplication&$\wedge$&$p$-form $\wedge$ $q$-form&$\rightarrow$&$ (p+q)$-form\\ 
Interior multiplication&$\rfloor$&vector $\rfloor$ $p$-form&$\rightarrow$&$ (p-1)$-form\\
Exterior derivative&$d$&$d$ $p$-form&$\rightarrow$&$ (p+1)$-form\\
Hodge star in a $n$-dimen. space&$^{\star}$&$^{\star} p$-form&$\rightarrow$&$ (n-p)$-form\\

\br
\end{tabular}
\end{indented}
\end{table}

\section{Units\label{UNITS_SECTION}}

In this work we made use of \textit{natural units}, i.e. $\hbar =c=1$
(cf table \ref{tabelle_6}).
\begin{table}
\caption{Natural units.}
\label{tabelle_6}
\begin{indented}
\item[]\begin{tabular}{@{}llll}
\br
[energy] & [mass] & [time] & [length] \\ 
\mr
length$^{-1}$ & length$^{-1}$ & length & length \\ 
\br
\end{tabular}
\end{indented}
\end{table}
Additionally, we have to be careful with the coupling constants and the
coordinates within the coframe. In order to keep things as clear as
possible, we provide a list of the quantities emerging throughout all sections in table \ref{tabelle_7}.
\begin{table}
\caption{Units of quantities.}
\label{tabelle_7}
\begin{indented}
\item[]\begin{tabular}{@{}lll}
\br
Quantities & \textit{1} & \textit{Length} \\ \mr
Gauge potentials & $[g_{\alpha \beta }],[\Gamma _{\alpha }{}^{\beta }]$ & $%
[\vartheta ^{\alpha }]$ \\ 
Gauge field strengths & $[Q_{\alpha \beta }],[R_{\alpha \beta }]$ & $[T^{\alpha }]$
\\ 
Gauge field excitations & $[M^{\alpha \beta }],[H^{\alpha }{}_{\beta }]$ & $%
[H_{\alpha }]^{-1}$ \\ 
Gauge field currents & $[E^{\alpha }{}_{\beta }],[m^{\alpha \beta }]$ & $%
[E_{\alpha }]^{-1}$ \\ 
Matter currents & $[\Delta _{\alpha \beta }],[\sigma ^{\alpha \beta }],[\tau
_{\alpha \beta }]$ & $[\Sigma _{\alpha }]^{-1}$ \\ \mr
Coordinates & $[\theta ],[\phi ],[r]$ & $[t]$ \\ 
Functions & $[\xi (t,r)],[\zeta (t)]$ & $[S(t)],[\mu (t)]^{-\frac{1}{4}%
},[p_{r}(t)]^{-\frac{1}{4}}$ \\ 
&  & $[p_{t}(t)]^{-%
\frac{1}{4}}$ \\
Miscellany &  & $[\Sigma _{\alpha \beta }]^{-\frac{1}{4}}$ \\ 
Constants & $[\chi ],[b],[k],[a_{I}]$ & $[\kappa ]^{\frac{1}{2}},[\Lambda
]^{-\frac{1}{2}},[c_{I}]^{-\frac{1}{2}},[c]^{-\frac{1}{2}}$ \\ 
&  & $[\Xi ]^{-\frac{1}{4}}$ \\ \mr
eq.\ (\ref{special_vacuum_solution_scale_factor_k<>0}) &  &
$[\varkappa_{1}]^{\frac{1}{2}},[\varkappa_{2}]$ \\ 
eq.\ (\ref{special_vacuum_solution_scale_factor_k=0}) &  & $[\varkappa
_{1}],[\varkappa _{2}]$ \\ 
eq. (\ref{scale_factor_lambda<>0_1}) &  & $[\varkappa _{1}],[\varkappa _{2}]$
\\ 
eq.\ (\ref{parameter_equation_of_state_1_ansatz}) & $[w]$ &  \\
eq.\ (\ref{parametrized_first_condition}) &  & $[\iota ]$ \\
eq.\ (\ref{solution_for_w_lambda<>0_case_ansatz_2}) &  & $[\varkappa _{3}]^{-%
\frac{1}{2}}$ \\\br
\end{tabular}
\end{indented}
\end{table}
Note that $[d]=1$ and $[\,^{\star }]=$ length$^{n-2p}$, where $n=$ dimension
of the spacetime, $p=$ degree of the differential form on which $^{\star }$
acts.

\ack
The author is grateful to F.W. Hehl and to the members of the gravity
group at the University of Cologne for their support. Additional thanks
go to an anonymous referee for useful suggestions concerning the
comprehensibility of this work.

\end{document}